\documentstyle[psfig]{mn}

\def\etal{et~al.}
\def\spose#1{\hbox to 0pt{#1\hss}}
\def\lta{\mathrel{\spose{\lower 3pt\hbox{$\mathchar"218$}}
     \raise 2.0pt\hbox{$\mathchar"13C$}}}
\def\gta{\mathrel{\spose{\lower 3pt\hbox{$\mathchar"218$}}
     \raise 2.0pt\hbox{$\mathchar"13E$}}}
\def\Ha{H$\alpha$}
\def\Hb{H$\beta$}

\def\kms{\,km\,s\,$^{-1}$}
\def\Ho50{$H_0 = 50$km\,s$^{-1}$\,Mpc$^{-1}$}

\title[Emission line gas of 3CR radio galaxies]{Ionisation, shocks and
evolution of the emission line gas of distant 3CR radio galaxies}

\author[P.~N.~Best \etal]{P.~N.~Best,$^1$\thanks{Email:
pbest@strw.leidenuniv.nl} H.~J.~A.~R\"ottgering$^1$ and
M.~S.~Longair$^2$\\ 
$^1$ Sterrewacht Leiden, Postbus 9513, 2300 RA Leiden, the Netherlands\\ 
$^2$ Cavendish Astrophysics, Madingley Road, Cambridge, CB3 0HE, UK} 

\pagerange{\pageref{firstpage}--\pageref{lastpage}}
\pubyear{1999}
\begin{document}
\label{firstpage}
\setcounter{topnumber}{3}
\setcounter{dbltopnumber}{3}

\maketitle

\begin{abstract}
\noindent An analysis of the kinematics and ionisation state of the
emission line gas of a sample of 14 3CR radio galaxies with redshifts $z
\sim 1$ is carried out. The data used for these studies, deep long--slit
spectroscopic exposures from the William Herschel Telescope, are presented
in an accompanying paper. It is found that radio sources with small linear
sizes ($\lta 150$\,kpc) have lower ionisation states, higher emission line
fluxes and broader line widths than larger radio sources. An analysis of
the low redshift sample of Baum \etal\ demonstrates that radio galaxies at
low redshift show similar evolution in their velocity structures and
emission line ratios from small to large radio sources.

The emission line ratios of small radio sources are in agreement with
theoretical shock ionisation predictions and their velocity profiles are
distorted. Together with the other emission line properties this indicates
that shocks associated with the radio source dominate the kinematics and
ionisation of the emission line gas during the period that the radio
source is expanding through the interstellar medium. Gas clouds are
accelerated by the shocks, giving rise to the irregular velocity
structures observed, whilst shock compression of emission line gas clouds
and the presence of the ionising photons associated with the shocks
combine to lower the ionisation state of the emission line gas. By
contrast, in larger sources the shock fronts have passed well beyond the
emission line regions; the emission line gas of these larger radio sources
has much more settled kinematical properties, indicative of rotation, and
emission line ratios consistent with the dominant source of ionising
photons being the active galactic nucleus.

This strong evolution with radio size of the emission line gas properties
of powerful radio galaxies mirrors the radio size evolution seen in the
nature of the optical--ultraviolet continuum emission of these sources,
implying that the continuum alignment effect is likely to be related to
the same radio source shocks.

\end{abstract}


\section{Introduction}

Powerful high redshift ($z \gta 1$) radio galaxies display a number of
remarkable characteristics. Their near infrared emission shows them to be
amongst the most massive galaxies known in the early Universe and to have
radial light profiles following de Vaucouleurs law, indicating that they
are fully formed giant elliptical galaxies \cite{bes98d}. At optical and
rest--frame ultraviolet (UV) wavelengths, however, the galaxies have very
irregular morphologies, frequently showing a strong excess of emission
aligned along the axis of the radio source \cite{mcc87,cha87}. Their
emission line properties are equally spectacular; luminous emission line
regions surround the radio galaxies, extending for several tens of
kiloparsecs or more (e.g. McCarthy \etal\ 1995)\nocite{mcc95}, with
velocity shears up to a few hundred \kms\ and line widths as high as
1500\kms\ (e.g. McCarthy \etal\ 1996)\nocite{mcc96a}. The emission line
regions are characterised by high ionisation spectra including strong
emission from species such as NeV and CIV. The origin of this luminous
emission line gas, its kinematics and ionisation, and their connection to
the radio source phenomenon remain important astrophysical questions.

Over the past few years we have been carrying out a detailed investigation
of a sample of 28 radio galaxies with redshifts $z \sim 1$ from the
revised 3CR catalogue \cite{lai83}, using optical imaging with the Hubble
Space Telescope (HST), high resolution radio interferometry with the Very
Large Array (VLA) and near infrared imaging with UKIRT
\cite{lon95,bes96a,bes97c,bes98d}. Here, and in an accompanying paper
(Best \etal\ 1999; hereafter Paper 1)\nocite{bes99b}, results are
presented from a deep spectroscopic campaign using the William Herschel
Telescope (WHT) on 14 of these radio galaxies, to study in detail the
emission line gas.  The reader is referred to Paper 1 for details of the
sample selection, data reduction, the reduced one dimensional spectra and
tabulated line fluxes, and the distributions of the intensity, velocity
and line width of the emission line gas as a function of position along
the slit for each galaxy.

The current paper is concerned with investigating the galaxy to galaxy
variations in the ionisation and kinematics of the emission line gas, and
comparing these variations with the radio and optical properties. The
layout is as follows. In Section~2 the photoionisation and shock
ionisation models are discussed and their predictions are compared with
the observed emission line ratios of the galaxies on an emission line
diagnostic diagram. The kinematics of the gas are studied in Section~3. In
Section~4 the results are compared with those of a low redshift sample of
galaxies and a scenario is proposed to explain the observed properties of
the emission line gas at low and high redshifts. The spectroscopic
properties of the distant galaxies are compared to their optical
properties, and the implications of these results for the continuum
alignment effect are discussed. Conclusions are drawn in
Section~5. Throughout the paper, values for the cosmological parameters of
$\Omega = 1$ and $H_0 = 50$\,km\,s$^{-1}$\,Mpc$^{-1}$ are assumed.

\section{The ionisation of the extended emission line regions}
\label{ionise}

Robinson \etal\ \shortcite{rob87} showed that for most low redshift ($z
\lta 0.1$) radio galaxies the emission line spectrum can be explained
adequately if it is assumed that the gas is photoionised by a power--law
emission source such as that provided by an active galactic nucleus
(AGN). Baum \etal\ \shortcite{bau92} obtained similar results for a sample
of radio galaxies out to redshift 0.2, although they noted that
photoionisation models could not reproduce [NII]~6584\,/\Ha\ ratios as
high as were observed for some galaxies; they suggested that localised
sources of heating and ionisation, for example shocks or the UV continuum
of surrounding hot gas (e.g. Heckman \etal\ 1989)\nocite{hec89}, may play
a role in some radio sources. McCarthy \shortcite{mcc93} constructed a
composite spectrum from a large sample of radio galaxies with redshifts
$0.1 < z < 3$; he showed that the emission line spectra of these more
distant (more radio powerful) sources are also consistent with being
photoionised, but as Villar--Mart{\'i}n \etal\ \shortcite{vil97} argued, this
composite spectrum is dominated by the few most highly ionised galaxies
and so this result does not necessarily apply to the population as a
whole. This photoionisation mechanism is in complete agreement with the
currently popular orientation--based unification schemes of radio galaxies
and radio loud quasars (e.g. Barthel 1989)\nocite{bar89} in which all
radio galaxies should host a powerful obscured active galactic nucleus,
supplying a large flux of anisotropically emitted ionising photons. The
presence of such an obscured quasar nucleus is also indicated by the
detection of spatially extended polarised emission and broad permitted
lines in polarised light, due to scattering of the AGN light by electrons
or dust (see Antonucci 1993 for a review)\nocite{ant93}.

Photoionisation is not the only story, however. As reviewed by Binette
\etal\ \shortcite{bin96}, simple photoionisation models fail to reproduce
some important features of the emission line spectra of the narrow line
regions of active galaxies. In particular, the strengths of many high
excitation lines (e.g. [NeV]~3426, CIV~1549, and high ionisation Fe lines)
are under--predicted by factors as large as 10, the electronic
temperatures derived by simple photoionisation models are too low when
compared with those inferred from the line ratio
[OIII]\,4363\,/\,[OIII]\,5007, and photoionisation models alone cannot
reproduce the large observed scatter in the HeII\,4686\,/\,\Hb\
ratio. Moreover, a significant fraction of radio galaxies show indications
of interactions between the radio jets and the surrounding emission line
gas, with the radio source shocks determining the morphology and
kinematics of the gas. Detailed studies of individual sources
(e.g. PKS\,2250$-$41, $z=0.31$, Clark \etal\ 1997, Villar--Mart{\'i}n
\etal\ 1999; 3C171, $z=0.24$, Clark \etal\ 1998)\nocite{cla97,cla98} have
shown that in some regions of the source the shocks can also dominate the
ionisation; for example, minima in the ionisation state are observed
coincident with the radio hotspots, and an anticorrelation is found
between the ionisation state of the extended gas and its (jet shock
broadened) line width.

At high redshifts, $z \gta 0.6$, interactions between the radio jets and the
gas are readily apparent from the kinematics of the ionised gas (see
Section~\ref{kinematics}). What has not been clear, however, is to what
extent the shocks play a role in the ionisation of these high redshift
sources.

\subsection{The CIII]~2326\,/\,CII]~1909 {\it vs} 
[NeIII]~3869\,/\,[NeV]~3426 diagram}
\label{CNediag}

Line ratio diagnostic diagrams, pioneered by Baldwin \etal\
\shortcite{bal81}, provide a powerful tool for investigating the
ionisation mechanism of emission line gas. They have been widely used to
distinguish the extended emission line regions of active nuclei from HII
regions and planetary nebulae, and in recent years also between shock and
photoionisation models for AGN.  Standard emission line diagnostics at
rest--frame optical wavelengths are shifted in to the near--infrared
wavebands for redshifts $z \gta 1$, and so cannot easily be used for high
redshift radio galaxies.  In the past couple of years, however, new
diagnostic diagrams have been constructed for rest--frame UV emission
lines redshifted into the optical window \cite{vil97,all98}.

The emission line pairs CIII]~2326 \& CII]~1909 and [NeIII]~3869 \&
[NeV]~3426 are well--suited for use as line ratio diagnostics for distant
radio galaxies for a number of reasons. All four of these fairly high
excitation lines are relatively strong in AGN spectra, and so their fluxes
can be determined with sufficient accuracy even for high redshift radio
galaxies. The two lines in each pair involve the same element, and
therefore there is no dependence of the line ratios on metallicity or
abundances; they are also close in wavelength, and so the effects of
differential reddening or calibration errors are minimised. Perhaps most
importantly, the predictions of shock and photoionisation models for these
line ratios are significantly different.

\begin{figure*}
\centerline{
\psfig{figure=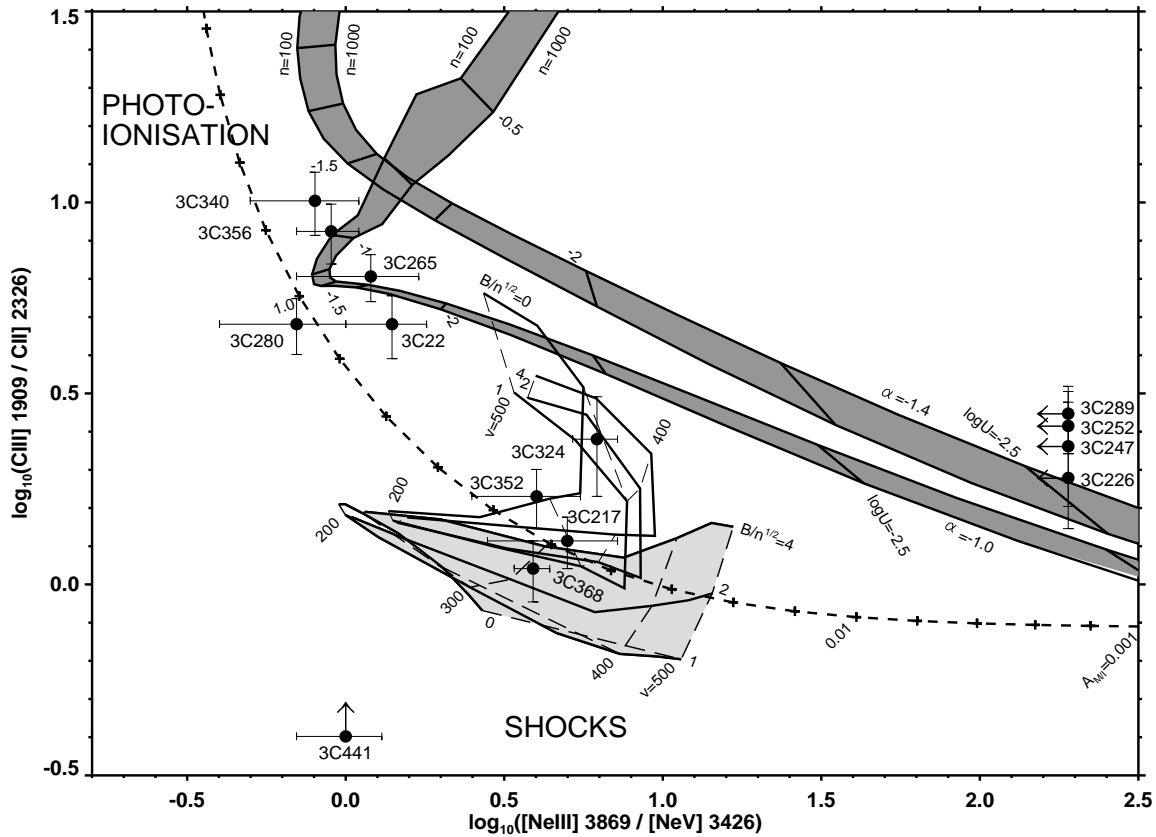,clip=,angle=90,width=15.8cm}
}
\caption{\label{linediag} An emission line diagnostic plot for the 3CR
radio galaxies, compared with theoretical predictions. The upper shaded
regions correspond to simple photoionisation models ($\alpha = -1.0$ and
$\alpha = -1.4$), as described in Section~\ref{photmods}. The dashed line
corresponds to the sequence for photoionisation models including matter
bounded clouds described in Section~\ref{matmods}. The lower shaded region
covers the ratios predicted by the shock models described in
Section~\ref{shockmods}; the unshaded region just above this corresponds
to the shock models including a precursor region (see
Section~\ref{shockmods}).  The five galaxies plotted towards the edge of
the diagram have no data available for one of their emission line ratios.
An interpretation of this diagram can be found in
Section~\ref{interpdiag}.}
\end{figure*}

The line ratio CIII]~2326\,/\,CII]~1909 was determined for 13 of the 14
sources in the sample from the data presented in Paper 1; the values of
this ratio are given in Table~\ref{props} together with many other derived
quantities of the radio galaxies. For the lowest redshift source, 3C340,
CII]~1909 is not redshifted to a sufficiently high wavelength to be
observed. The ratio [NeIII]~3869\,/\,[NeV]~3426 is only available from the
Paper 1 data for the source 3C324, but for nine of the other sources in
the sample values are available from the literature; these also are
compiled in Table~\ref{props}. The two line ratios are plotted against
each other on Figure~\ref{linediag}, where they are compared to the
theoretical predictions of shock and photoionisation as discussed in the
following subsections.

\begin{table*}
\caption{\label{props} Ionisation and kinematic properties of the radio
galaxies, as calculated from the data presented in Paper 1. Column 1 gives
the radio source name and column 2 its redshift. The projected linear size
of the radio source (from Best \etal\ 1997) is given in column 3. The
emission line ratios of CIII]~2326\,/\,CII]~1909 and
[NeIII]~3869\,/\,[NeV]~3426, together with their uncertainties, are given
in columns 4 to 7. The integrated [OII]~3727 flux density is given in
column 8 and its equivalent width in the rest--frame of the source is in
column 9. The projected linear extent of the emission line region along
the direction of the slit is in column 10. Column 11 gives the maximum
FWHM of the [OII]~3727 emission line in the spectrum, column 12 the range
in relative velocities of this line seen along the slit, and column 13 the
number of distinct velocity components, as described in the text. Column 14 
gives the extent of the aligned optical emission from the HST images
of Best \etal\ (1997).}
\begin{center}
\begin{tabular}{lccccccccccccc}
Source & z & Radio & \multicolumn{2}{c}{CIII]\,/\,CII]}
&\multicolumn{2}{c}{[NeIII]\,/\,[NeV]$^a$} &[OII] flux & Eq. & Emis. & Max. 
& Vel. & No. & Opt. \\
&& Size & Ratio & Error & Ratio & Error
&($\times 10^{-15}$) & width & Size & FWHM & Range & Comps & Size \\
&& [kpc]&&&&&[erg/s/cm$^{2}$]&[\AA]&[kpc] &[km/s]&[km/s]&[kpc]\\ 
\multicolumn{1}{c}{(1)}&(2)&(3)&(4)&(5)&(6)&(7)&(8)&(9)&(10)&(11)&(12)&(13)&(14) \\
3C22   & 0.935 & 208 & 4.8 & 0.9 & 1.4 & 0.4 & 2.02 & 80  &  41 & 875  & 75  & 1 & 16 \\
3C217  & 0.898 & 110 & 1.3 & 0.2 & 5.0 & 2.2 & 4.85 & 544 &  57 & 1000 & 175 & 3 & 21 \\
3C226  & 0.818 & 263 & 1.9 & 0.3 & --- & --- & 1.21 & 159 &  50 & 725  & 425 & 1 & 19 \\
3C247  & 0.749 & 113 & 2.3 & 0.9 & --- & --- & 1.24 & 129 &  77 & 850  & 250 & 3 & 43 \\
3C252  & 1.104 & 488 & 2.6 & 0.4 & --- & --- & 0.81 & 125 &  56 & 450  & 250 & 1 & 22 \\
3C265  & 0.810 & 646 & 6.4 & 0.9 & 1.2 & 0.5 & 3.56 & 191 & 111 & 700  & 725 & 4 & 76 \\
3C280  & 0.997 & 117 & 4.8 & 0.8 & 0.7 & 0.3 & 2.11 & 173 & 102 & 800  &650&$4\pm1$&22\\
3C289  & 0.967 & 89  & 2.8 & 0.5 & --- & --- & 0.78 & 149 &  46 & 675  &  75 & 1 & 16 \\
3C324  & 1.208 & 96  & 2.4 & 0.7 & 6.2 & 1.0 & 1.93 & 219 &  60 & 1025 &800&\,2$^b$&25\\
3C340  & 0.775 & 371 & 10.1& 1.9 & 0.8 & 0.3 & 0.58 & 84  &  47 & 600  & 200 & 1 & 24 \\
3C352  & 0.806 & 102 & 1.7 & 0.3 & 4.0 & 1.5 & 2.57 & 295 &  78 & 1050 & 800 & 3 & 24 \\
3C356$^c$&1.079& 624 & 8.4 & 1.5 & 0.9 & 0.2 & 0.76 & 111 &  32 & 725  &  50 & 1 & 12 \\
3C368  & 1.132 & 73  & 1.1 & 0.2 & 3.9 & 0.5 & 5.87 &\,202$^d$&86&1350 &600&$3\pm1$&58\\
3C441  & 0.708 & 211 & --- & --- & 1.0 & 0.3 & 0.47 &  58 &  33 & 900  & 375 & 1 & 18 \\
\end{tabular}
\end{center}
\raggedright Notes:\\
\raggedright [a] Values of the [NeIII]~3869\,/\,[NeV]~3426 ratio are taken
from: 3C22 --- Rawlings \etal\ (1995); 3C217 \& 3C340 --- Spinrad, private
communication;  3C265 \& 3C352 --- Smith \etal\ (1979); 3C280 --- Spinrad
(1982); 3C324 --- Paper 1;  3C356 --- Lacy \& Rawlings (1994); 3C368 ---
Stockton \etal\ (1996); 3C441 --- Lacy \etal\ (1998).\\
\raggedright [b] For 3C324 this corresponds to the two kinematically
distinct components at +400 and $-$400\kms.\\ 
\raggedright [c] The data for 3C356 are for more northerly galaxy.\\
\raggedright [d] For 3C368 the determined equivalent width is a lower
limit due to the contribution of the M-star (Hammer et~al 1991) to the
continuum level.
\end{table*}
\nocite{raw95,smi79,spi82,lac94,lac98,sto96a,bes98d,ham91}

\subsection{Photoionisation models}
\label{photmods}

The theoretical line ratios of CIII]~2326\,/\,CII]~1909 and
[NeIII]~3869\,/\,[NeV]~3426 for photoionised gas were taken from the work
of Allen \etal\ \shortcite{all98} who calculated the predicted line ratios
for a number of emission lines using the MAPPINGS II code \cite{sut93}.
Allen \etal\ considered a planar slab of gas illuminated by a power-law
spectrum of ionising radiation, and calculated the emission line ratios
for a wide range of conditions: for two different spectral indices of the
input spectrum ($F_{\nu} \propto \nu^\alpha$ with $\alpha = -1$ and
$\alpha = -1.4$), and two different densities of cloud ($n_{\rm e} = 100$
and 1000\,cm$^{-3}$), the ionisation parameter $U$\footnote{The ionisation
parameter $U$ is defined as the ratio of the number density of ionising
photons striking the cloud to the gas density ($n_{\rm H}$) at the front
face of the cloud $[U = (c n_{\rm H})^{-1} \int_{\nu_0}^{\infty} (F_{\nu}
{\rm d}\nu)/h\nu]$, where c is the speed of light and $\nu_0$ is the
ionisation potential of hydrogen.} was allowed to vary in the range
$10^{-4} \le U \le 1$. A high energy cutoff was applied to the ionising
spectrum at 1.36\,keV to avoid over-producing the intensity of the soft
X-rays. The models are ionisation bounded and correspond to a range in
cloud sizes from 0.003 to 32 parsec. The resultant line ratio sequences
are shown on Figure~\ref{linediag}.

It is beyond the scope of this paper to provide a more detailed
description of the photoionisation models, or of the other theoretical
models considered next. For more complete discussions of these
modelling techniques the reader is referred to the papers from which
the theoretical line ratios have been drawn, in this case the work of
Allen \etal\ \shortcite{all98}. 

\subsection{Photoionisation including matter bounded clouds}
\label{matmods}

To avoid the shortcomings of simple photoionisation models discussed at
the beginning of Section~\ref{ionise}, Binette \etal\ \shortcite{bin96}
considered photoionisation of a composite population containing both
optically thin (matter bounded; MB) and optically thick (ionisation
bounded; IB) clouds (cf. Viegas and Prieto 1992).\nocite{vie92} In their
models, all of the photoionising radiation passes initially through the MB
clouds which absorb a fraction ($F_{\rm MB} \sim 40$\%) of the impinging
ionising photons and produce the majority of the high ionisation lines in
the spectrum. The radiation which isn't absorbed then strikes the
population of IB clouds; this radiation has already been filtered by the
MB clouds as a result of which the IB clouds give rise to predominantly
low and intermediate excitation lines. According to these models, the
variation in the emission line ratios from galaxy to galaxy has its origin
in the variation of the ratio (hereafter $A_{\rm M/I}$) of the solid angle
from the photoionising source subtended by MB clouds relative to that of
IB clouds. A larger value of $A_{\rm M/I}$ corresponds to a larger weight
given to the MB clouds and hence a higher excitation spectrum. Since the
model states that all of the ionising radiation striking the IB clouds
must first have passed through the MB clouds, the ratio $A_{\rm M/I}$
strictly cannot be below unity.

Binette \etal\ consider two physical situations which might produce such a
composite cloud population (some combination of the two would also be
possible): (i) the MB `clouds' may be optically thin shells surrounding a
denser IB core of a cloud; (ii) the MB clouds could be a separate
population of clouds which lie close to the ionising source. In the second
case, the MB clouds would have to have a covering factor of unity in order
that all lines of sight to the IB clouds pass through a MB cloud; this is
possible for very small clouds, consistent with the fact that they would
be optically thin. Note that in the case of the MB clouds forming a
separate cloud population, if some fraction of them are obscured from the
observer, for example by the same material that obscures the active
nucleus itself, this may give rise to an apparent $A_{\rm M/I} < 1$.

Binette \etal\ tabulated the line ratios in the two cloud populations for
a single set of parameters, chosen to give a good match to Seyfert
spectra. They adopted a power-law spectrum with a spectral index of
$\alpha =-1.3$ ($F_{\nu} \propto \nu^\alpha$), ionisation parameters of
$U_{\rm MB}=0.04$ and $U_{\rm IB} = 5.2 \times 10^{-4}$, a density in the
MB clouds of 50\,cm$^{-3}$, and absorbed ionising photon fractions of
$F_{\rm MB} = 0.4$ and $F_{\rm IB} = 0.97$ (see their paper for a more
detailed discussion of these quantities). From these data, a sequence of
line ratios of CIII]~2326\,/\,CII]~1909 and [NeIII]~3869\,/\,[NeV]~3426
have been calculated allowing the quantity $A_{\rm M/I}$ to vary in the
range $0.001 \le A_{\rm M/I} \le 100$; this sequence is shown on
Figure~\ref{linediag}.

\subsection{Shock ionisation models}
\label{shockmods}

Fast radiative shocks are a powerful source of ionising photons which can
have a profound influence upon the temperature and ionisation properties
of the gas in the post--shock region. An overview of shock ionisation
models is provided by Dopita \& Sutherland \shortcite{dop96}. The two most
important parameters for controlling the post--shock emission line
spectrum are the velocity of the shock and the ratio $B/\sqrt {n}$, where
$B$ is the pre--shock transverse magnetic field and $n$ is the pre--shock
number density of the emission line clouds. The latter ratio controls the
density, and hence the effective ionisation parameter, of the post--shock
gas since at high shock velocities the transverse magnetic field limits
the compression caused by the shock through a balance between the magnetic
pressure of the cloud ($\propto B^2$) and the ram pressure of the shock
($\propto n$).

Dopita and Sutherland \shortcite{dop96} calculated the emission line
ratios expected from gas ionised by the photons produced in shocks for a
range of physical conditions: the velocity of the shock through the
emission line clouds was allowed to vary from 150 to 500\kms, and the
`magnetic parameter' was varied in the range $0 \le B/\sqrt {n} \le
4$\,$\mu$G\,cm$^{-1.5}$, which spans the expected range of values (see
their paper for more details). These authors also emphasised the importance
of photons produced by the shock diffusing upstream and ionising the
pre--shock gas. This may give rise to extensive precursor emission line
regions, with different spectral characteristics to the compressed shocked
gas. They therefore calculated the emission line spectra predicted for
these precursor regions for a range of shock velocities from 200 to
500\kms. In distant radio galaxies such as the ones studied in this paper,
the spatial resolution is insufficient to distinguish between the
precursor and post-shock emission regions, and a combined spectrum of the
two would be observed.

Using the data tables provided by Dopita \& Sutherland, the emission line
ratios of CIII]~2326\,/\,CII]~1909 and [NeIII]~3869\,/\,[NeV]~3426 were
calculated both for simple shock models and for shock models including a
precursor region. These theoretical ratios are shown on
Figure~\ref{linediag}.

\subsection{Interpreting the line diagnostic diagram}
\label{interpdiag}

\begin{figure}
\centerline{
\psfig{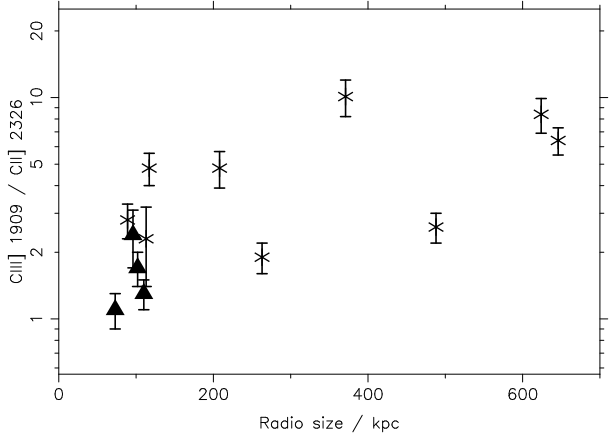} 
}
\caption{\label{cratsize} The correlation between the CIII]~2326 /
CII]~1909 emission line ratio and the projected linear size of the radio
source. The four sources lying in the `shocks' region of the line
diagnostic diagram, Figure~\ref{linediag}, are plotted as filled triangles
and the remainder of the galaxies as asterisks.}
\end{figure}

Figure~\ref{linediag} clearly demonstrates that the ionisation states of
radio galaxies, even within a tightly defined sample such as the one
studied here, show considerable variations: the CIII]~2326\,/\,CII]~1909
ratio differs by nearly a factor of ten between 3C368 and 3C340. The nine
sources for which ratios are available for both lines clump into two
groups of sources; 3C217, 3C324, 3C352 and 3C368 are grouped together in
the region corresponding to the shock ionisation models, while 3C22,
3C265, 3C280, 3C340 and 3C356 are all close to the photoionisation
predictions. Interestingly, the four sources in the first region all have
projected radio linear sizes smaller than 115\,kpc and the five sources in
the second group all have sizes larger than this value. This result is
more apparent in Figure~\ref{cratsize} where the ratio of
CIII]~2326\,/\,CII]~1909\footnote{This ratio is preferred to the
[NeIII]~3869\,/\,[NeV]~3426 ratio here, and in later figures, because data
are available for more of the galaxies and because the ratios are all
drawn from the homogeneous data set presented in Paper 1. The Neon ratio
provides similar results, as is apparent from the strong inverse
correlation between the two ratios seen on Figure~\ref{linediag}.} is
plotted against the linear size of the radio source; the two parameters
are correlated at the 98.5\% significance level in a Spearmann--Rank test.

For all nine sources, the photoionisation models of Binette \etal\
\shortcite{bin96} including matter--bounded clouds provide an acceptable
fit to the data. There is, however, a problem with this model. As
discussed in Section~\ref{matmods}, a value of $A_{\rm M/I} < 1$ is only
possible if the MB clouds form a separate population of clouds, some of
which are obscured from the observer. The four small radio sources would
have to have a value $A_{\rm M/I} \approx 0.1$, implying that over 90\% of
their MB clouds are obscured, while there is no requirement for
obscuration of the larger radio sources ($A_{\rm M/I} \approx 1$). The
amount of obscuration would therefore have to depend upon the size of the
radio source. 

Three factors determine the projected linear size of a radio source: the
orientation of the source with respect to the line of sight, the age of
the radio source and the advance rate of the hotspots. The first of these
cannot be responsible since sources orientated more towards the line of
sight, hence appearing smaller, will have less, not more, obscuration
towards their central regions (cf. orientation--based schemes of radio
galaxies and radio loud quasars). Although the second option cannot be
excluded, it seems improbable that the obscuration of MB clouds in the
central regions will decrease by an order of magnitude during the short
timescale of a radio source lifetime (a few $\times 10^7$\,years) without
the same process either destroying the clouds themselves or having other
consequences, for example for the visibility of the broad--line
regions. The third possibility, that there may be a connection between a
higher obscuration of the central regions of small radio sources and a
slower advance rate of their hotspots, has parallels with the suggestion
that compact symmetric radio sources are small because they are confined
by a dense (obscuring) surrounding medium (e.g. van Breugel et~al
1984)\nocite{bre84}. However, recent investigations of compact radio
sources using VLBI techniques have derived hotspot advance velocities of
significant fractions of the speed of light ($\approx 0.2c$, e.g. Owsianik
\& Conway 1998, Owsianik et~al 1998)\nocite{ows98a,ows98b}, supporting a
youth rather than confinement scenario for these sources. This indicates
that there is no strong connection between hotspot advance speed and
obscuration (density) of the central regions of radio sources.

It seems unlikely, therefore, that the value of $A_{\rm M/I}$ should be
strongly dependent upon the radio source size. Although this possibility
cannot categorically be excluded, our preferred interpretation of the line
diagnostic diagram is that for four sources, for which the extent of the
emission line region is within a factor of two of the radio source size,
the ionisation is dominated by shocks, whilst for the other five sources
photoionisation dominates. This interpretation will be supported by the
discussion of the kinematics in the following section; the matter--bounded
cloud photoionisation model would provide no clear explanation of the
variations seen in the kinematical properties.

The uncertainties in the emission line ratios for each individual galaxy
are too large to pin down any parameters of the ionisation accurately, but
one feature is readily apparent. The five sources in the photoionisation
region of the diagram are significantly more consistent with a flatter
spectral index ($\alpha \approx -1.0$) for the power--law ionising
continuum than with the steeper one ($\alpha \approx -1.4$) typically
adopted for low redshift sources. Villar--Mart{\'i}n \etal\ \shortcite{vil97}
found a similar result analysing the rest--frame UV emission lines of
radio galaxies with redshifts $z > 1.7$, suggesting that this may be a
general feature of high redshift AGN.

\section{Morphological and kinematical properties of the emission line gas}
\label{kinematics}

The emission line gas surrounding low redshift radio galaxies shows
velocity shears within the galaxies of between 50 and 500\kms, and
(deconvolved) full width at half maxima (FWHM) of the emission lines
typically in the range 200 to 600\kms \cite{tad89b,bau92}. In many cases
the kinematics are consistent with a gravitational origin. At high
redshifts the kinematics can be much more extreme, with velocity
dispersions often in excess of 1000\kms\ (McCarthy \etal\ 1996, Paper
1)\nocite{mcc96a,bes99b} and components offset by several hundreds of
\kms\ with respect to bulk of the gas (Tadhunter 1991, Paper
1)\nocite{tad91,bes99b}. These remarkable kinematics are inconsistent with
gravitational origins (cf. Tadhunter 1991)\nocite{tad91}. In this section
the variation in the kinematics is compared with other properties of the
radio source to investigate their origin.

In Table~\ref{props} a number of the parameters of the emission line
properties of the gas are provided. The integrated [OII]~3727 emission
line intensity and the rest--frame equivalent width of this emission line
are as calculated in Paper 1. The projected linear size of the emission
line region along the slit direction was determined from the extent of the
locations at which fits to the [OII]~3727 emission line profile were
obtained in Figures~2 to 15(d) of Paper 1 (excluding the detached emission
line systems for 3C356 and 3C441). The range in relative velocities was
calculated, to the nearest 25\kms, from Figures~2 to 15(e) of Paper 1,
considering the velocity separation between the most positive and most
negative velocity components of the [OII]~3727 emission line, excluding
any data points with uncertainties greater than 100\kms. The maximum value
of the FWHM of the emission line gas was determined from Figures~2 to
15(f) in Paper 1, again excluding any locations with uncertainties greater
than 100\kms.

One further parameter was calculated, hereafter referred to as the `number
of velocity components', $N_{\rm v}$, to provide an indication of the
smoothness of the velocity profile. $N_{\rm v}$ was defined as the number
of single velocity gradient components (ie. straight lines) necessary to
fit, within the errors, the velocity profiles along the slit direction
(Figures~2 to 15e of Paper 1; cf van Ojik \etal\ 1997).\nocite{oji97} A
galaxy whose mean motion is consistent with simple rotation will provide a
single component fit; higher values of $N_{\rm v}$ correspond to irregular
motions. This analysis, being by its very nature somewhat subjective, was
carried out separately by two of the authors and by a third independent
scientist. For 11 of the 14 galaxies a unanimous value of $N_{\rm v}$ was
obtained. The remaining three galaxies show more complicated profiles and
their classification is ambiguous: for 3C280, values of 3, 4 and 5 were
obtained, and so a value of $4 \pm 1$ is adopted; for 3C324 the profile is
very different from the other galaxies, being composed of two
kinematically distinct systems as discussed in Paper 1 --- a value of 2 is
used; values of 2, 3 and 4 were assigned to 3C368, and so $3 \pm 1$ is
adopted. The precise values of $N_{\rm v}$ for these galaxies are of less
importance than the fact they they are clearly inconsistent with a value
of 1. The values of $N_{\rm v}$ are compiled in Table~\ref{props}.

\begin{figure}
\centerline{
\psfig{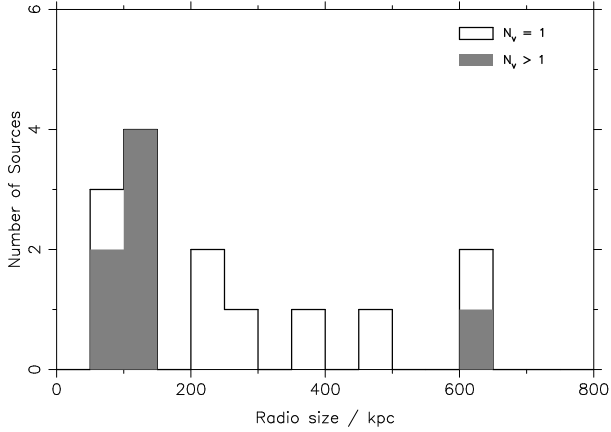}
}
\caption{\label{nvfig} A histogram of the radio size distribution of the
sources, separated into sources with smooth velocity profiles ($N_{\rm v}
= 1$, unshaded) and those whose profiles are irregular ($N_{\rm v} > 1$,
shaded).}
\end{figure}

A number of features are immediately apparent from Table~\ref{props}. It
is noteworthy that of the seven galaxies with projected radio sizes
smaller than 150\,kpc, six have values $N_{\rm v} > 1$ with only one
having $N_{\rm v} = 1$, while six of the seven sources larger than this
size have $N_{\rm v} = 1$ (see Figure~\ref{nvfig}). A chi--squared test
shows that the probability of this occurring by chance is below 1\%. Small
radio sources predominantly have emission line gas with distorted velocity
profiles and the emission line gas of large radio sources has a velocity
profile generally consistent with rotation.

\begin{figure}
\centerline{
\psfig{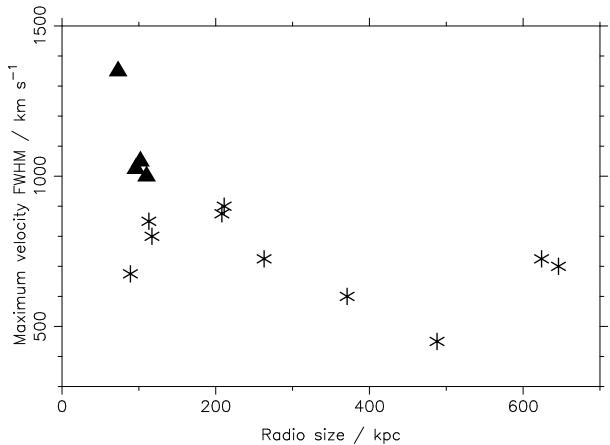}
}
\caption{\label{fwhmsize} The inverse correlation between the maximum FWHM
of the [OII]~3727 emission line and the projected linear size of the radio
source. Symbols as in Figure~\ref{cratsize}. }
\end{figure}

\begin{figure}
\centerline{
\psfig{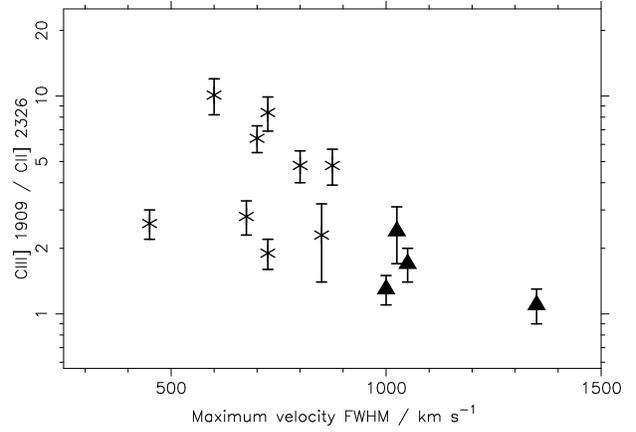}
}
\caption{\label{cratfwhm} A plot showing the direct connection between the
ionisation state of the emission line gas, as indicated by the
CIII]~2326\,/\,CII]~1909 line ratio, and its kinematics in terms of the
emission line maximum FWHM. Symbols as in Figure~\ref{cratsize}.}
\end{figure}

A similar result is found with the variation of the maximum FWHM with
radio size, shown in Figure~\ref{fwhmsize}. These two parameters are
anti-correlated at greater than the 99\% significance level (Spearmann
Rank test), with the four sources lying in the `shock' region of the line
diagnostic diagram (Figure~\ref{linediag}) having clearly the highest
values. This latter point is made more clearly in Figure~\ref{cratfwhm}
where the FWHM of the [OII]~3727 emission can be seen to be inversely
correlated with the CIII]~2326\,/\,CII]~1909 emission line ratio, at the
98.5\% significance level using a Spearmann Rank correlation test. The
kinematical and ionisation properties of these galaxies are fundamentally
connected.

\begin{figure}
\centerline{
\psfig{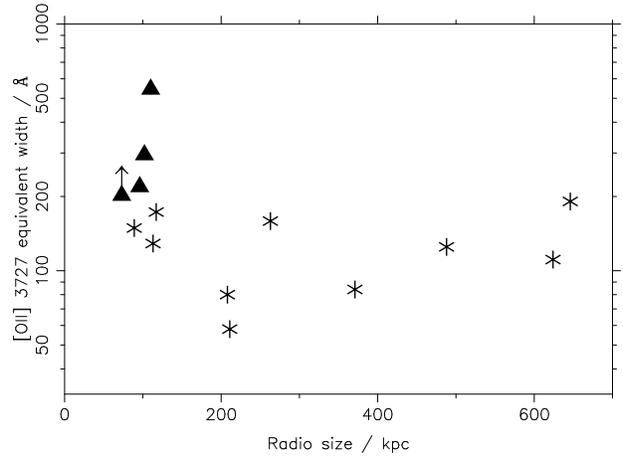}
}
\caption{\label{EWsize} The decrease in the equivalent width of the
[OII]~3727 emission line with increasing size of the radio source. Symbols
as in Figure~\ref{cratsize}. 3C368 is plotted as a lower limit due to the
contribution of the M-star to its continuum level.}
\end{figure}

It is not only the kinematics of the gas that evolve with the radio source
size, but also the physical extent and the luminosity of the line
emission. Figure~\ref{EWsize} shows the variation of the equivalent width
of the [OII]~3727 emission line with increasing size of the radio
source. Although this correlation is less strong (96\% significance in a
Spearmann Rank test), it is apparent that the small sources in the
`shock--dominated' region of Figure~\ref{linediag} show enhanced
[OII]~3727 equivalent widths. A more accurate description of
Figure~\ref{EWsize} is not that there is an inverse correlation between
the equivalent width of the [OII] emission and radio size, but rather that 
at large ($\gta 150$\,kpc) radio sizes the distribution of equivalent 
widths is fairly flat, and at small sizes there
is often a factor of 2 to 3 excess emission relative to this level.

\begin{figure}
\centerline{
\psfig{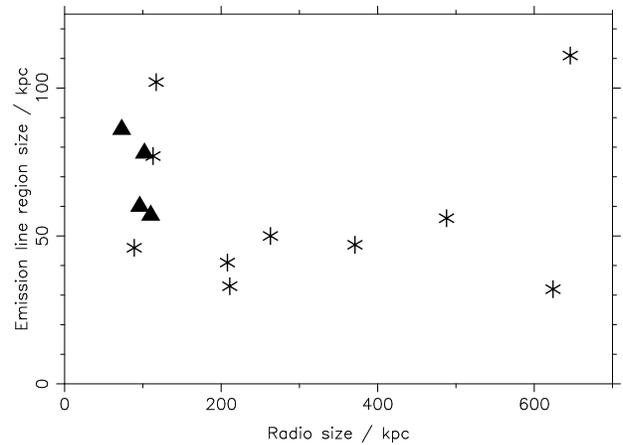}
}
\caption{\label{emissize} The variation of the linear extent of the
[OII]~3727 emission line region with the size of the radio source. Symbols
as in Figure~\ref{cratsize}.}
\end{figure}

This enhancement of the line {\em equivalent widths} of small radio
sources with respect to large sources implies an even greater boosting of
their line {\em luminosities}, for two reasons. First, the optical
continuum emission of small radio sources is more luminous than that of
large sources, as indicated by Best \etal\ \shortcite{bes96a}, decreasing
the apparent increase in the emission line equivalent width. Second, the
equivalent width is determined from the extracted 1--dimensional spectrum
from a spatial region along the slit of about 35\,kpc (see Paper 1); the
physical extent of the emission line regions of small radio sources is
greater than that of large radio sources, as shown in
Figure~\ref{emissize}. Excluding 3C265, which is an exceptional source in
many ways (e.g. see discussion in Best \etal\ 1997), the emission line
regions of radio sources with sizes $\gta 200$\,kpc have total extents of
up to about 50\,kpc (25\,kpc radius, if symmetrical). Smaller radio
sources, however, have emission line regions ranging from this size up to
about 100\,kpc, a size comparable to the extent of the radio source. In
other words, line emission at distances from the AGN of 30 to 50\,kpc
generally is only seen at the stage of radio source evolution when the
hotspots are passing, or have just passed, through this region.

\section{Discussion}

A number of results have been derived in the previous sections and these
are summarised here for clarity.

\begin{itemize}
\item Radio sources with small linear sizes ($\lta 120$\,kpc) have lower
ionisation states than larger radio sources. Their emission line ratios
are in agreement with the theoretical predictions of shock ionisation
models, whilst those of large radio sources are consistent with
photoionisation. 

\item There is a strong inverse correlation between the FWHM of the
[OII]~3727 emission and the size of the radio source. The four sources
with `shock--dominated' ionisation states have the highest FWHM.

\item Large radio sources often have smooth velocity profiles consistent
with rotation, whilst those of small sources are more distorted.

\item The [OII]~3727 emission line strength correlates inversely with the
radio source size. The four `shock--dominated' sources have the highest
integrated [OII]~3727 equivalent widths.

\item The physical extent of the line emitting regions is larger in smaller 
radio sources.
\end{itemize}

Before discussing the interpretation of these correlations, first a
comparison is made to see if such results also hold for low redshift radio
galaxies.

\subsection{Comparison with low redshift radio galaxies}

Baum \etal\ \shortcite{bau92} studied the ionisation and kinematics of a
sample of 40 radio galaxies with redshifts $z \lta 0.2$. Their sample
contained a large mixture of radio source types, including both Fanaroff
\& Riley (1974; hereafter FR)\nocite{fan74} class I and II
objects\footnote{FR\,I radio sources are edge--darkened sources of
generally lower radio luminosity that the FR\,II sources; FR\,II's are
characterised by bright hotspots towards the extremities of each lobe.} as
well as sources with intermediate structures. Many differences are now
known to exist between the FR\,I and FR\,II sources besides the large
differences at radio wavelengths, such as the luminosity and
environments of their host galaxies \cite{hil91,bau95,led96}, the
luminosity of their emission line gas \cite{zir95}, differences in the
dust properties \cite{dek99}, and possibly a different mode of accretion
on to the central black hole \cite{rey96}.

Baum \etal\ also found a significant difference in the host galaxy
kinematics between the two radio source types. They classified the
kinematics of the radio galaxies into three classes, `rotators', `calm
non--rotators' and `violent non--rotators'. They found that almost all of
the FR\,II sources fell into the rotator or violent non-rotator classes;
most of the FR\,I and intermediate type sources were calm
non-rotators. All of the FR\,II's had strong emission lines with a
relatively high ionisation parameter, whilst the FR\,I and intermediate
class sources had much weaker emission lines of lower ionisation, with the
surrounding hot interstellar and intracluster medium likely to play an
important role in the ionisation, both through heat conduction and through
ionisation by its ultraviolet and soft X--ray emission (see also Baum
\etal\ 1995, Zirbel \etal\ 1995)\nocite{bau95,zir95}. Given the large
differences between FR\,I and FR\,II sources, to allow a direct comparison
with the high redshift sample, attention here is restricted to only the
FR\,II sources in their sample. Four FRII's in the sample lie more
southerly than declination $-30^{\circ}$ and for two of these accurate
determinations of the radio size could not be found in the literature; to
avoid introducing any biases by selecting only the well-studied sources,
all four of these sources have been excluded from further consideration.

\begin{table}
\caption{\label{baumtab} Selected properties of the low redshift sample of
FR\,II radio sources studied by Baum \etal\ (1992).}
\begin{center}
\begin{tabular}{lcccc}
Source & Radio size & Kinematic &\multicolumn{2}{c}{[OI]~6300.3\,/\,\Ha}\\
              & [kpc] &Class$^a$ & Ratio  & Error \\
3C33          &  397  &   R   &  0.20  &  0.08 \\
3C63          &  84   &  VNR  &  0.28  & \,---$^b$ \\
3C98          &  264  &   R   &  0.12  &  0.01 \\
PKS\,0634-206 & 1200  &   R   &  0.06  &  0.01 \\
3C192         &  308  &   R   &  0.10  & \,---$^b$ \\
3C227         &  499  &   R   &  ---   &  ---  \\
3C285         &  268  &   R   &  0.15  &  0.09 \\
3C403         &  357  &   R   &  0.21  &  0.16 \\
3C405         &  194  &  CNR  &  0.32  &  0.08 \\
3C433         &  169  &  VNR  &  0.23  &  0.03 \\
PKS\,0349-278 &  635  &   R   &  ---   &  ---  \\
3C171         &  171  &  VNR  &  ---   &  ---  \\
3C184.1       &  473  &   R   &  ---   &  ---  \\
3C277.3       &  106  &  VNR  &  ---   &  ---  \\
4C29.30       &   85  &  VNR  &  ---   &  ---  \\
3C293         &  261  &   R   &  ---   &  ---  \\
3C305         &   15  &   R   &  ---   &  ---  \\
3C382         &  287  &   R   &  ---   &  ---  \\ 
\end{tabular}
\end{center}
\raggedright Notes:\\
\raggedright [a] R --- rotator; CNR --- calm non--rotator; VNR --- violent
non--rotator.\\ 
\raggedright [b] No error quoted as the mean value is taken from data at
only one position.
\end{table}

\begin{figure}
\centerline{
\psfig{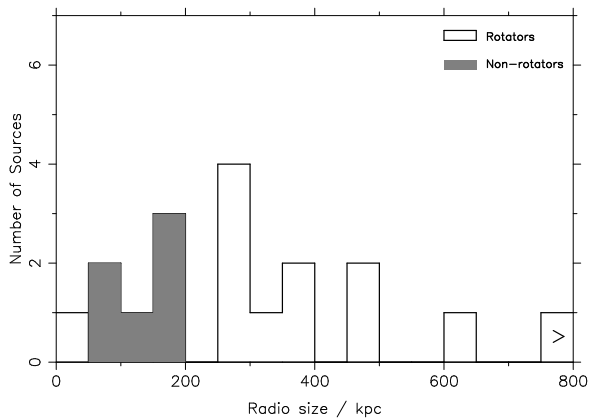}
}
\caption{\label{baumfig1} A histogram of radio sizes for the different
kinematic classifications of the FR\,II radio galaxies in the low redshift
Baum \etal\ sample.}
\end{figure}

The remaining sample of low redshift FR\,II radio galaxies is listed in
Table~\ref{baumtab}, along with the linear size of the radio source taken
from the literature and the kinematic classification given by Baum \etal\
\shortcite{bau92}. In Figure~\ref{baumfig1} a histogram of the linear
sizes is presented, separating the rotator and non-rotator classes. It is
clear that the non-rotator classes are associated with small radio
sources, and the rotator class with larger sources, exactly as is found
for the high redshift sample. The only exception to this rule is 3C305,
which is a rotator with a small radio size: indeed, this is the smallest
radio source in the sample (15\,kpc), and it could be argued that any
shocks associated with the radio source have not yet passed through a
significant proportion of the host galaxy, accounting for the lack of
clear non-rotational kinematics. Even including 3C305, the probability of
the radio sizes of the rotator and non-rotator classes being drawn from
the same parent samples is less than 0.5\% (using a Mann--Whitney
U--test).

\begin{figure}
\centerline{
\psfig{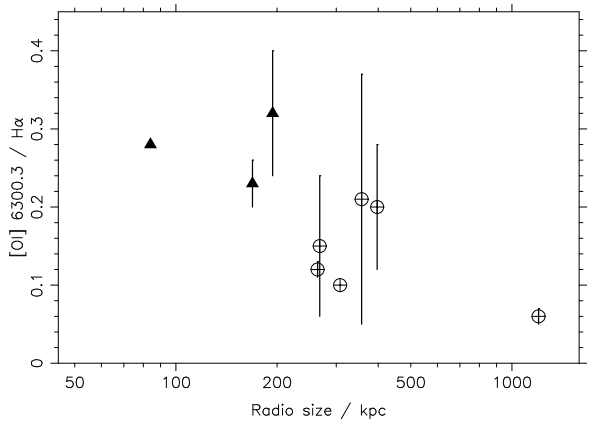}
}
\caption{\label{baumfig2} A plot showing the anticorrelation between
the radio size and the ionisation state for the FR\,II's in the low
redshift Baum \etal\ sample. The rotators are represented by
the crossed circles and the non-rotator by the filled triangles.}
\end{figure}

All of the FR\,II's in the sample have relatively high ionisation states,
but differences are seen from galaxy to galaxy. Baum \etal\
\shortcite{bau90} present the line strengths of the [OI]~6300.3,
[NII]~6548.1,6583.4, \Ha\ and [SII]~6716.4,6730.8 emission lines as a
function of position for half of the sample considered in their 1992
paper. Although these are all relatively low ionisation lines and
therefore not the most sensitive to differences between shock and
photoionisation, the [OI]\,/\,\Ha\ ratio should be somewhat higher for
shock ionised gas than for photoionised gas. An `average' value of this
emission line ratio has been calculated for each galaxy as the mean
of the ratios at the various positions tabulated by Baum \etal\
\shortcite{bau90}; these are given in Table~\ref{baumtab}, the errors
quoted representing the scatter in the ratio with location in the
galaxy. In Figure~\ref{baumfig2} these ratios are plotted against radio
size: a Spearmann rank test shows that this emission line ratio is
anticorrelated with radio size at the 96\% confidence level.

The low redshift sample therefore provides similar results to the high
redshift sample. Large FR\,II radio sources have kinematics consistent
with rotation and higher ionisation states than small radio sources, whose
ionisation and kinematics show more evidence for the role of shocks. It is
of note that the low and intermediate redshift sources for which
individual studies have shown unambiguously that the kinematics and
ionisation are dominated by shocks are almost invariably cases in which
the radio source is of comparable size to the extended emission line
regions (e.g. Clark \etal\ 1997,1998)\nocite{cla97,cla98}, naturally
agreeing with this picture.

One significant difference that remains between the low and high redshift
samples is that the high redshift sources are more extreme in their
emission line properties (luminosities, line widths, etc) than those at
low redshifts. The most important factor influencing this is the sharp
increase of the radio power with redshift in the flux--limited samples,
with corresponding increases in both the flux of ionising photons from the
AGN and the energy of the jet shocks. However, Tadhunter \etal\
\shortcite{tad98} investigated the correlations of different emission line
strengths with redshift and showed that this cannot be the only reason:
the ionisation--sensitive [OII]~3727 / [OIII]~5007 ratio does not decrease
strongly with redshift as it should if the only difference between the low
power, low redshift and the high power, high redshift objects was that the
latter contained a more luminous photoionising source. They concluded that
a secondary effect such as an increase in the density of the intergalactic
medium or an increase in the importance of jet-cloud interactions with
redshift is also required. 

\subsection{The role of shocks in small sources}
\label{origemis}

The results presented in the previous sections, coupled with the
evidence that a similar situation is seen at low redshifts, lead
naturally to a single scenario to explain all of the emission line
properties.

For small radio sources the morphology, kinematics and ionisation
properties of the emission line gas are dominated by the effects of the
bow shock associated with the expansion of the radio source. As this bow
shock passes through the interstellar and intergalactic medium (ISM \&
IGM), the inter--cloud gas is quickly accelerated to the velocity of the
bow shock, but the warm emission line clouds are essentially bypassed by
the shock front (e.g. Rees 1989, Begelman \& Cioffi
1989)\nocite{ree89,beg89}.  The clouds are accelerated during the short
time it takes the shock to pass the cloud by, due to the imbalance in the
pressures between the pre--shock and post--shock gas on the front and back
of the cloud. The velocity to which the clouds are accelerated in this way
is easily shown to be independent of cloud size and to be well below
100\,kms (e.g. Rees 1989).\nocite{ree89}

Much larger velocities are induced, however, if the effect of
ram--pressure acceleration by the shocked IGM gas (often referred to as
entrainment) is considered. Behind the initial bow shock, the clouds find
themselves in a shocked layer of IGM, moving outwards at speeds
approaching that of the bow shock. The clouds will be accelerated within
this medium until they pass across the contact discontinuity into the
radio cocoon, where the pressure is the same as in the shocked layer of
gas but the density is much lower, and they are no longer accelerated;
there is essentially no mixing of the hot inter--cloud gas across this
contact discontinuity (e.g. Norman \etal\ 1982)\nocite{nor82}.

During the time $\Delta T$ for which a cloud is between the bow shock and
the contact discontinuity, the momentum imparted to the cloud by the
shocked IGM can be approximated to first order as $r_{\rm c}^2 v_{\rm s}^2
n_{\rm g} m_{\rm p} \Delta T$, where $r_{\rm c}$ is the cloud size,
$v_{\rm s}$ is the bow shock velocity, $n_{\rm g}$ is the post--shock
number density of the inter--cloud gas and $m_{\rm p}$ is the proton
mass. The mass of the cloud is of order $r_{\rm c}^3 n_{\rm c} m_{\rm p}$,
where $n_{\rm c}$ is the cloud number density, and the timescale $\Delta
T$ is of order $D / v_{\rm s}$, where $D$ is the distance between the
bow--shock and the contact discontinuity; therefore, the velocity to which
the cloud is accelerated is

\begin{equation}
\label{equn}
v_{\rm c} \sim \frac{D}{r_{\rm c}} \frac{n_{\rm g}}{n_{\rm c}} v_{\rm s}.
\end{equation}

In the radio source evolution models of Kaiser \& Alexander
\shortcite{kai97a}, radio sources grow self--similarly and $D$ is found to
be about 3\% of the distance between the AGN and the bow shock
\cite{kai99a}; for the radio source passing through the emission line
region at radius $\sim 15$\,kpc then, $D \sim 0.5$\,kpc. Assuming a
density ratio of $n_{\rm g} / n_{\rm c} \sim 10^{-4}$ for pressure
equilibrium between the clouds ($T \sim 10^4$\,K) and the surrounding IGM
($T \sim 10^8$\,K), and a shock velocity of $v_{\rm s} \sim 0.05c$ from
typical hot-spot advance velocities (e.g. Liu, Pooley \& Riley
1992)\nocite{liu92}, then a cloud of size $r_{\rm c} \sim 1$\,pc will be
accelerated to 750\kms, comparable to the velocities observed in the small
radio sources (the actual velocities may need to be slightly higher since
the radio galaxies are believed to lie close to the plane of the sky). The
spread in projected cloud velocities from clouds of different sizes and
from clouds accelerated through different regions of the bow shock will
lead to the broad velocity dispersions. The acceleration of the emission
line gas clouds by the radio bow shocks therefore explains the distorted
velocity profiles and large line widths observed in small radio sources.

It is interesting to note that the acquired cloud velocities are
proportional to the bow--shock velocity $v_{\rm s}$. If the bow--shock
velocity increases with radio power (redshift), as has been suggested from
spectral ageing measurements of hotspot advance velocities (e.g. Liu,
Pooley \& Riley 1992)\nocite{liu92}, this would explain why greater
velocity widths are seen in high redshift sources than low redshift
sources.

The ionisation state of the large radio sources indicates that the
dominant source of ionising photons in these sources is the AGN. Since the
properties of the AGN are not expected to change dramatically between
small and large radio sources, the gas surrounding the small sources
should receive a similar flux of photoionising radiation. The lower
ionisation state seen in the spectra of these galaxies arises in part due
to compression of emission line gas clouds by the radio source shocks,
decreasing the ionisation parameter. The presence of extra (softer)
ionising photons associated with the shocks further influences the
ionisation state. Bicknell \etal\ (1997; see also Dopita and Sutherland
1996)\nocite{bic97,dop96} have investigated the emission line luminosity
that can be generated by radio source shocks expanding through a
single--phase ISM with a power--law density gradient, $\rho(r) = \rho_{\rm
0} (r/r_{\rm 0})^{-\delta}$. For $\delta = 2$, they show that the work
done on the ISM by the expanding radio cocoon ($P{\rm d}V$) is
approximately half of the energy supplied by the radio jet; if the shock
is fully radiative then a significant proportion of this energy is fed
into emission line luminosity. The luminosity of the [OIII]~5007 emission
line can be estimated as

\begin{displaymath}
L({\rm [OIII]}) \approx \frac{6}{8-\delta} \left(\frac{\kappa_{\rm
1.4}}{10^{-11}}\right)^{-1} \left( \frac{P_{1.4}} {10^{27}{\rm
W\,Hz^{-1}}} \right) \times 10^{43} {\rm ergs\,s^{-1}},
\end{displaymath} 

\noindent where $P_{1.4}$ is the monochromatic power of the radio source
at 1.4\,GHz, and $\kappa_{\rm 1.4}$ is the conversion factor from the
energy flux of the jet to the monochromatic radio power at 1.4\,GHz, which
Bicknell \etal\ \shortcite{bic97} estimate to be of order
$10^{-10.5}$. Adopting this value, taking the flux density of a typical $z
\sim 1$ 3CR source at an observed frequency of 1.4\,GHz to be 2\,Jy, and
assuming that the [OII]~3727\,/\,[OIII]~5007 emission line flux ratio is
$\sim 0.5$ \cite{mcc93}, then for $\delta = 2$ the observed [OII]~3727
emission line flux produced by the shocks is calculated to be $f({\rm
[OII]}) \sim 3 \times 10^{-15}$\,ergs\,s$^{-1}$\,cm$^{-2}$. Of this,
probably between a third and a half (that is, 1 to $1.5 \times
10^{-15}$\,ergs\,s$^{-1}$\,cm$^{-2}$) will fall within the projected sky
area from which the spectrum was extracted. This predicted emission line
flux can be compared to the [OII]~3727 emission line fluxes observed in
the data (Table~\ref{props}), which lie in the range $\sim 0.5$ to $5
\times 10^{-15}$\,ergs\,s$^{-1}$\,cm$^{-2}$ with the smaller radio sources
typically having the higher values (see also Figure~\ref{EWsize}). These
results are completely consistent with a small (factors of 2 to 5) boosting
of the emission line luminosities of small sources due to the extra energy
input from the shocks.

Once the radio source shocks have passed beyond the emission line clouds,
the shock--induced emission line luminosity will fall. Under the simplest
assumptions, once the jets pass beyond the confining ISM the pressure
inside the cocoon will drain away, and the cocoon wall shocks will no
longer be pressure driven (e.g. Dopita 1999)\nocite{dop99}. These shocks
will pass into a momentum conserving phase; their velocity will decrease
roughly as $v_{\rm s} \propto r^{-2}$, and so since the shock luminosity
per unit area scales as $v_{\rm s}^3$, the shock--induced emission line
luminosity will fall as $r^{-4}$. Although these assumptions are
oversimplified, taking no account of confinement by an intracluster medium
for example, it is clear that once the shock fronts have now passed well
beyond the emission line regions, the contribution of ionising photons
produced by the shocks will decrease rapidly; this is in complete accord
with the larger sources having photoionisation dominated emission line
regions.

\subsection{The physical extent of the emission line gas}
\label{gasext}

The physical extent of the emission line region of each galaxy along the
slit direction was provided in Table~\ref{props}. For comparison, the
extent of the aligned optical (rest--frame ultraviolet) emission has also
been determined from the HST observations of Best \etal\
\shortcite{bes97c}; using the HST image taken through the filter at a
rest--frame wavelength of about 4000\AA, the angular distance over which
optical emission was observed at greater than three times the rms sky
noise level of the image was measured for each galaxy, and the
corresponding `optical sizes' are given in Table~\ref{props}. These values
can further be compared with these results of Best \etal\
\shortcite{bes98d}, who showed from near--infrared imaging that, underlying
the aligned emission, the radio sources are hosted by giant elliptical
galaxies with characteristic radii of typically 10 to 15\,kpc.

The extent of the optical aligned emission does not exceed 25\,kpc except
in three cases: 3C247, 3C265 and 3C368. For 3C368, the HST `continuum'
image is actually dominated by a combination of line emission and the
correspondingly luminous nebular continuum emission (see discussion in
Section~\ref{contal}). The large extent of 3C247 is also likely to be
predominantly line emission, since it arises from a diffuse halo of
emission exactly tracking that seen in a narrow--band [OII]~3727 image by
McCarthy \etal\ \shortcite{mcc95}. 

With the exception of 3C265 (which, as discussed in Paper 1, is an unusual
source in many ways), it is therefore reasonable to say that the aligned
continuum emission has an extent of only a couple of characteristic radii,
and so lies within the body of the host galaxy. The situation with the
emission line gas is very different: this has a physical extent which can
exceed 100\,kpc, with a mean extent of over 60\,kpc. The emission line gas
clearly extends well beyond the confines of the host galaxy. As was shown
in Figure~\ref{emissize}, there is also a difference in the physical
extent of the line emitting regions between large and small radio sources,
with line emission at radii of 30 to 50\,kpc generally only seen in small
radio sources. Unless there is an intrinsic difference between the
environments of the small and large radio sources, which seems unlikely
given all of the correlations found, the emission line gas clouds must
also be present out to radii $\gta 30$\,kpc in large radio sources, but is
not visible.

Again, the role of shocks can be considered to explain this. At these
radii, the flux of ionising photons from the active nucleus may be
insufficient to produce an observable emission line luminosity. As the
radio source shocks pass through these regions, however, the gas density
will be increased and, as discussed above, a large source of local
ionising photons will become available, pushing up the emission line
luminosity. Following the passage of the radio shocks and the consequent
removal of the associated ionising photons, this enhanced line emission
will fade over timescales much shorter than the radio source
lifetime. Thus, luminous line emission is only seen from the clouds at
radii 30 to 50\,kpc at the time that the radio source shocks are passing
through these regions. A direct consequence of this model is that for
radio sources smaller than about 100\,kpc a positive correlation between
radio source size and emission line region size should be observed, since
line emission from the clouds at radii 30 to 50\,kpc will not be seen
until the radio source has advanced that far. Such a correlation has
indeed been observed in the Ly-$\alpha$ emission of radio galaxies with $z
\ga 2$ \cite{oji97}.

An interesting test of the model presented here could be carried out by
taking high spatial resolution long--slit spectra of a sample of radio
galaxies with radio sizes smaller than the size of the emission line
regions. The prediction is that within the region of the host galaxy
occupied by the radio source, the radio source shocks will be important;
the emission line ratios will be consistent with shock ionisation, and the
gas kinematics will be distorted with broad velocity dispersions. Outside
of this region, however, the gas clouds will not yet have been influenced
by the radio source shocks and photoionisation should dominate. A study
with a similar principle has been carried out on the radio source
1243+036, a radio galaxy of radio size about 50\,kpc at a redshift $z =
3.6$. Distorted Ly-$\alpha$ velocity structures with large velocity FWHM
are seen within the radio source structure, but Ly-$\alpha$ emission also
extends beyond that to at least 75\,kpc radius in an apparently rotating
halo \cite{oji96a}.  Villar--Mart{\'i}n \etal\ \shortcite{vil99a} have
also found that the line emission of PKS\,2250$-$41 ($z=0.308$) is
composed of distinct kinematic components: a low ionisation component with
broad velocity width in the region of the radio source structure, and a
narrower high ionisation component which extends beyond the radio lobe.
Carrying out studies such as these for a large sample of radio sources is
important because the velocity structures of the line emission in regions
outside the radio shocks will directly show the initial motions of the
emission line clouds and can be used to determine whether these clouds are
simply material associated with the formation of the galaxy which has been
expelled into the IGM, or whether they have an external origin, brought in
either by a galaxy merger or a cooling flow. It is difficult to
distinguish between such scenarios in larger radio sources since
information on the initial cloud velocities has been destroyed by the bow
shock acceleration.

\subsection{Evolution of the velocity structures}
\label{velevol}

One significant issue remains to be explained in this picture, and that is
how the velocity structures of the large radio sources are produced. The
high gas velocities and velocity dispersions induced by the shocks in
small radio sources are seen to evolve within the timescale of a radio
source lifetime, a few $\times 10^7$ years, such that the emission line
clouds obtain an underlying velocity profile consistent with a rotating
halo, albeit with a still high velocity dispersion. Questions that need to
be considered are whether this truly is rotation that is being seen, over
what timescale can the extreme shock--induced kinematics be damped down,
and can a mean rotation profile be produced whilst the velocity dispersion
remains so high?

Regarding the first question, given the single slit position and
relatively low spatial resolution for the high redshift radio galaxies, it
cannot categorically be stated that the emission line profiles of large
radio sources are rotation profiles. The data are consistent, for example,
with outflow along the radio axes, although in this case it is not clear
why the velocity increases with radius (a structure more like that of
3C324 -- see Paper 1 -- might be expected) while the velocity dispersion
decreases with increasing source size. At low redshifts, however, much
higher spatial resolution studies using multiple slit positions show
clearly that the gas is in rotating structures (e.g. Baum \etal\
1990)\nocite{bau90}. It therefore seems reasonable to assume that this may
also be true at higher redshifts, and even if this is not the case, the
questions noted above still need to be addressed for the low redshift
radio sources.

Three plausible mechanisms can be considered for the evolution in the
velocities of the emission line clouds over the radio source lifetime. The
first is that the emission line clouds settle back into stable orbits
within the host galaxy through gravitational dynamics alone. The timescale
for this process is of order a few crossing times of the clouds, where for
clouds moving with velocity $v_{\rm c} \sim 500$\kms\ at a radius $r \sim
15$\,kpc in the galaxy, the crossing time is $t_{\rm c}
\sim 2r/v \sim 6 \times 10^7$\,years. This timescale is longer than the
radio source lifetime, and so gravity alone cannot give rise to the
observed evolution in the emission line structures.

A second possibility concerns the deceleration of emission line clouds
moving with respect to the interstellar medium, due to ram--pressure
arguments. This works through the same process as the acceleration
argument discussed in Section~\ref{origemis}.  As the emission line clouds
move through the inter--cloud gas, those clouds moving with the largest
velocities sweep up the greatest mass of inter--cloud gas, and so are
decelerated most strongly. This process will decrease the width of the
cloud velocity distribution.

Simulations have been carried out, as detailed in Appendix~\ref{append},
to investigate the timescale over which the mean velocity of an
ensemble of emission line clouds (with an initial velocity distribution
similar to that seen in small radio sources) evolves to that of the IGM in
which the clouds are moving, and the timescale over which the dispersion
of the velocity distribution is decreased. It is found that the peak of
the velocity distribution evolves to that of the gas in which it is moving
considerably more quickly than the velocity width decreases. Both
timescales depends upon the typical cloud size and the ratio of the cloud
density to that of the inter--cloud medium within the radio cocoon, and
for reasonable assumptions the timescale for decrease of the velocity
widths is comparable to the radio source lifetime (see
Appendix~\ref{append} for details).

Therefore, if a population of emission line clouds were placed within the
rotating ISM of a galaxy, a mean rotation profile for the emission line
clouds could be recovered whilst the FWHM of the emission lines remained
large, as is observed in the radio galaxies. The problem with this model,
however, is that the radio bow shock sweeps up essentially all of the
inter--cloud gas, with little mixing through the contact discontinuity
(e.g. Norman \etal\ 1982)\nocite{nor82}. The radio cocoon is filled
primarily with material supplied by the radio jets, and so there is
essentially no gas left following a rotation profile. Such gas would have
to be resupplied to the ISM, for example by supernovae and stellar winds
from stars in rotational orbits, but it is very unlikely that enough gas
can be supplied in this manner. Alternatively, the cocoon material
supplied by the radio jets would itself have to be in rotational motion,
perhaps through angular momentum transfer from the rotating IGM to the
radio source as the bow shocks advance. To summarise, although this
mechanism can decrease the velocity widths of the gas, it is not clear
whether a rotation profile can be re-established quickly enough.

The third possibility considers the evolution of the population of
radiating clouds through a combination of galaxy rotation and cloud
shredding. Klein \etal\ \shortcite{kle94} showed that in the aftermath of
a bow shock, emission line clouds may be susceptible to shredding due to
growing Kelvin--Helmholtz and Rayleigh--Taylor instabilities on their
surface. The clouds could be shredded over a timescale of a few `cloud
crushing times', $t_{\rm cc} \sim \chi^{1/2} r_{\rm c} / v_{\rm b}$, where
$\chi$ is the density ratio of the cloud to the surrounding medium inside
the cocoon, $r_{\rm c}$ is the post--shock cloud radius, and $v_{\rm b}$
is the velocity of the bow shock through the IGM. Kaiser \etal\
\shortcite{kai99b} considered such cloud disruption as a way to resupply
material to the radio cocoon in order to explain how a secondary hotspot
can be formed in the newly discovered class of double--double radio
galaxies (e.g. Schoenmakers \etal\ 1999)\nocite{sch99}; they derived a
value of $t_{\rm cc} \sim 5 \times 10^6 (r_{\rm c} / {\rm pc})$ yrs.

The cloud shredding time is shortest for the smallest emission line
clouds; clouds smaller than about a parsec will be shredded on timescales
shorter than the radio source lifetime. These small clouds were the most
rapidly accelerated (see Equation~\ref{equn}) and so are responsible for
producing much of the high velocity dispersions and distorted velocity
structures. If these high velocity small clouds are destroyed then the line
emission will become dominated by the remaining more massive clouds, which
were less accelerated by the radio source shocks, have a lower velocity
dispersion, and may still maintain the vestiges of a rotation profile.

Equation~\ref{equn} further shows that the velocity acquired by the warm
clouds is proportional to the velocity of the bow shock. In directions
perpendicular to the radio axis the bow shock velocity is lower by a
factor of the aspect ratio of the cocoon (typically between about 1.5 and
6, e.g. Leahy \etal\ 1989)\nocite{lea89}, and so the warm clouds in these
directions will be less accelerated. In small radio sources these clouds
will not be very luminous since they lie away from the strongest radio
source shocks and outside of the cone of photoionising radiation from the
partially obscured AGN; the emission will be dominated by the higher
velocity clouds along the radio jet direction. Over a rotation timescale
($\sim 10^7$ years), however, these low velocity clouds may be brought
within the ionisation cone of the AGN, become ionised, and contribute
significantly to the emission line luminosity. Likewise, clouds in small
radius orbits around the AGN will acquire lower velocities, since the
distance between the bow shock and the contact discontinuity is less and
so the period of acceleration is shorter. Thus the rotation profile may
re-establish itself from the central regions of the galaxy outwards.

By these two mechanisms of shredding and mixing of the cloud populations,
the observed population of emission line clouds will evolve such that, in
large radio sources, an increasing percentage of the emission will arise
from clouds which were less accelerated by the bow shock and so the
rotation profile will be gradually recovered. The cloud shredding model
has a further advantage that if some fraction of the clouds are destroyed
in large radio sources then the emission line luminosity will decrease
with increasing radio size, as is observed.  The one drawback of this
model is that it is surprising that the distinction between radio sources
showing rotation profiles and those with distorted profiles is so
sharp. Note also that if this scenario is the correct one then the
emission line clouds must lie in rotating orbits prior to the radio
source activity, providing some information as to their origin.

In conclusion, the observation of emission line clouds in rotating halos
around large radio galaxies is not trivial to explain, given the large
influence of the radio source bow shocks passing through the
medium. Gravitational effects alone cannot be responsible for
re--establishing rotation profiles, but a combination of cloud shredding
and cloud mixing, maybe with some help from ram--pressure deceleration,
could reproduce the effect.

\subsection{Implications for the alignment effect}
\label{contal}

In 1987, McCarthy \etal\ and Chambers \etal\ \nocite{mcc87,cha87}
demonstrated that the optical--UV emission of radio galaxies with
redshifts $z \gta 0.6$ has a strong tendency to be elongated and aligned
along the direction of the radio source. HST images of a sample of 28 of
these radio galaxies \cite{bes97c} have demonstrated that the form of this
so--called `alignment effect' varies strongly from galaxy to galaxy, and
in particular appears to evolve with increasing size of the radio source
\cite{bes96a}. Small radio sources show a number of intense blue knots
tightly aligned along the direction of the radio jet, whilst larger
sources generally have more diffuse optical--UV morphologies.  Given the
strong similarity between this radio size evolution of the continuum
alignment effect and the evolution of the emission line gas properties, it
is instructive to examine the role of the radio source shocks and the
emission line clouds in giving rise to continuum emission.

One direct connection is the nebular continuum emission from the warm
emission line gas clouds \cite{dic95}, that is, free--free emission,
free--bound recombination, two--photon continuum and the Balmer forest
lines. The flux density of this emission is directly connected to the flux 
of the \Hb\ emission line. The very luminous line emission seen
in the spectra of these powerful radio galaxies (e.g. Paper 1) thus
implies that nebular continuum emission is likely to make a significant
contribution to their UV flux density. Indeed, 3C368 was one of the
original three radio galaxies studied by Dickson \etal\ \shortcite{dic95},
and they found a nebular continuum contribution in the northern knots as
high as 60\% of the total continuum emission at rest--frame wavelengths
just below the 3646\AA\ Balmer break (see also Stockton \etal\
1996)\nocite{sto96a}. As can be seen from Table~\ref{props}, 3C368 is a
somewhat extreme case and the contribution for more typical galaxies will
be somewhat lower, but still of great significance.  In
Section~\ref{kinematics} it was shown that the luminosity of the emission
lines correlated inversely with the size of the radio source
(Figure~\ref{EWsize}); therefore, the strength of nebular continuum
emission will decrease with increasing radio source size, and in small
sources will be found predominantly along the radio jet tracing the
strongest radio source shocks. This reflects exactly the observed
evolution of the continuum alignment effect.

A second alignment effect hypothesis involving the emission line clouds is
that star formation is induced by the passage of the radio jet, due to the
radio source shocks compressing gas clouds and pushing them over the
Jean's limit (e.g. Rees 1989, Begelman and Cioffi 1989, De Young
1989)\nocite{ree89,beg89,dey89}. It should be noted that it is the most
massive clouds which would collapse to form the stars, and these are
distinct from the smallest clouds which are the most likely to be
destroyed by the bow shock. In regions which might be
star--forming, $\lta 10^6$\,years behind the bow shock, the only clouds
which will already have been destroyed by instabilities on their surface
are those of size $r_{\rm c} \lta 0.1$\,pc (see Section~4.3); for a mean
cloud density of 100\,cm$^{-3}$ this corresponds to a total cloud mass of
less than 10$^{-2} M_{\odot}$, not massive enough to have formed a star 
anyway.

As discussed by Best \etal\ \shortcite{bes96a}, the jet--induced star
formation mechanism can also account directly for the evolution of the
optical--UV morphology with radio size: the mass of stars required to
produce the excess optical--UV emission is only a few $\times 10^8
M_{\odot}$ \cite{lil84a,dun89a}, well below 1\% of the stellar mass of the
galaxy, and since the starburst luminosity drops rapidly with age they
become indistinguishable from the evolved star population over a timescale
of a few $\times 10^7$ years. On the negative side, no direct evidence for
young stars in these radio galaxies was found in our spectra (cf. 4C41.17
at higher redshift, $z=3.8$; Dey \etal\ 1997)\nocite{dey97}, although the
clearest features of young stellar populations fall outside the observed
wavelength ranges. 

Another important continuum alignment model is scattering of light from a
hidden quasar nucleus by electrons \cite{fab89a} or dust (e.g Tadhunter
\etal\ 1989a, di Serego Alighieri \etal\ 1989)\nocite{tad89a,dis89}.
Strong support for this model comes from the observation that the
optical emission of some distant radio galaxies is polarised at the $\sim
10$\% level with the electric vector oriented perpendicular to the radio
axis (e.g. Cimatti \etal\ 1996 and references therein)\nocite{cim96}, and
the detection of broad permitted lines in polarised light
\cite{dey96,cim96,tra98}: clearly some fraction of the excess optical--UV
emission must be associated with this mechanism. However, the lack of
polarised emission from some sources (e.g. 3C368, van Breugel \etal\ 1996;
see also Tadhunter \etal\ 1997)\nocite{bre96c,tad97} dictates that this is
not universal; even for 3C324 where the polarisation percentage is high,
only a fraction $\lta 30-50$\% of the optical--UV emission is associated
with the scattered component \cite{cim96}. A problem for scattering models
is that, in the simplest picture, a biconical emission region is expected
for the scattered light, rather than the knotty strings of emission
observed to lie along the radio jet. However, in light of jet--shock
models, this could be explained by extra scattering particles being made
available along the radio jet axis, either as dust grains being produced
in jet--induced star forming regions, or by radio source shocks disrupting
optically thick clouds along the radio jet direction and exposing
previously hidden dust grains \cite{bre96b}.

In conclusion, radio source shocks will play a key role in producing the 
observed morphology and radio size evolution of the continuum alignment
effect. Nebular continuum emission will be enhanced in small
radio sources, some gas clouds may be induced to collapse and form stars,
and extra scattering particles associated either with any star formation
or the disruption of gas clouds could enhance the scattered component.

\section{Conclusions}

The main conclusions of this work can be summarised as follows:

\begin{itemize}
\item Small radio sources show a lower ionisation state than large radio
sources. The emission line ratios of radio sources with linear sizes $\lta
120$\,kpc are consistent with the gas being ionised by photons produced by
the shocks associated with the radio source. The emission line
luminosities of the small sources are boosted by a small factor ($\sim
2-5$) relative to large sources, in accord with them receiving an extra
source of ionising photons from the shock.

\item Small radio sources have very distorted velocity profiles, large
velocity widths, and emission line regions covering a larger spatial
extent than those of large sources; the latter have much smoother velocity
profiles which appear to be dominated by gravitation. These properties are
fully explained in terms of the passage of the shocks associated with the
radio source.

\item A strong correlation is found between the ionisation state of the
gas and its kinematical properties, indicating that the two are
fundamentally connected.

\item These correlations, originally derived for the sample of redshift
one radio galaxies studied in Paper 1, are shown also to hold for a sample
of FR\,II radio galaxies with redshifts $z \lta 0.2$.

\item The similarity of the evolution of the emission line gas properties
with radio size to that of the continuum alignment effect makes a strong
case for the continuum alignment effect also having a large dependence
upon radio source shocks.

\item The continuum alignment effect is generally confined to within the
extent of the host galaxy, but line emission is observed over a
considerably larger spatial extent.

\end{itemize}

\section*{Acknowledgements} 

The William Herschel Telescope is operated on the island of La Palma by
the Isaac Newton Group in the Spanish Observatorio del Roches de los
Muchachos of the Instituto de Astrofisica de Canarias. This work was
supported in part by the Formation and Evolution of Galaxies network set
up by the European Commission under contract ERB FMRX-- CT96--086 of its
TMR programme. We are grateful to Mark Allen for supplying the output of
the MAPPINGS II photoionisation models in digitised form, and Hy Spinrad
for providing the Neon line ratios for 3C217 and 3C340. We thank Matt
Lehnert, Arno Schoenmakers and Christian Kaiser for useful discussions,
and the referee, Mike Dopita, for his careful consideration of the
original manuscript and a number of useful suggestions.

\bibliography{pnb} 
\bibliographystyle{mn}

\appendix

\section{Deceleration of emission line clouds moving through the IGM} 
\label{append}

Consider a spherical cloud of emission line gas with number density
$n_{\rm c}$ and radius $r_{\rm c}$ travelling at velocity $v_{\rm c}$
through gas of number density $n_{\rm g}$ and velocity $v_{\rm g}$. In a
time ${\rm d}t$ a mass of gas of approximately $\pi 
r_{\rm c}^2 n_{\rm g} m_p (v_{\rm c}-v_{\rm g}) {\rm d}t$, where $m_p$ is
the proton mass, is displaced by the cloud and accelerated from velocity
$v_{\rm g}$ to velocity $v_{\rm c}$. The momentum of the cloud is
correspondingly decreased:

\begin{displaymath}
\frac{4}{3} \pi r_{\rm c}^3 n_{\rm c} m_p {\rm d}v_{\rm c} = \pi r_{\rm
c}^2 n_{\rm g} m_p (v_{\rm c}-v_{\rm g})^2 {\rm d}t 
\end{displaymath}

Defining $t_0$ as $t_0 = 4 n_{\rm c} r_0 / 3 n_{\rm g}$, where $r_0$ is
a typical cloud radius, then

\begin{displaymath}
{\rm d}v_{\rm c} = \frac{(v_{\rm c} - v_{\rm g})^2}{r_{\rm c}/r_0}
\frac{{\rm d}t}{t_0} 
\end{displaymath}

\begin{figure}
\psfig{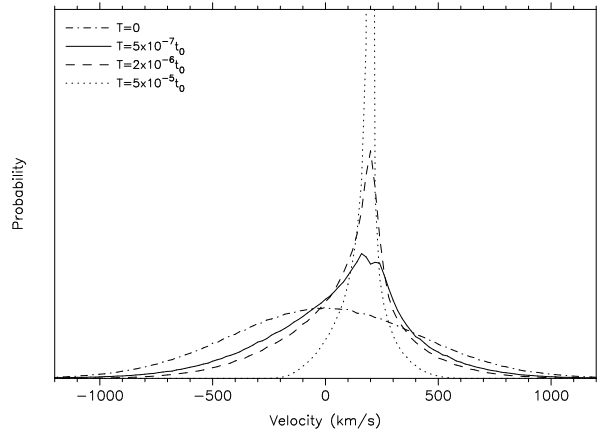}
\caption{\label{monte} Modelling the evolution of the emission line cloud
velocity distribution within gas moving at 200\kms.}
\end{figure}

Using this equation it is possible to follow the evolution of an ensemble
of such emission line clouds. For simplicity the distribution of emission
line cloud radii was chosen to be flat in logarithm space over a factor of
1000 range centred on $r_0$, that is, $\rm{P}(\rm {log}(r_{\rm c}))$ is
constant in the range $-1.5 \le {\rm log}(r_{\rm c}/r_0) \le 1.5$, and 0
outside that range. The initial velocity distribution of the clouds was
set to follow a Gaussian distribution with a mean velocity of zero and a
FWHM of 1000\kms, chosen to represent the velocity dispersion observed in
small radio sources. A Monte Carlo simulation was then used to follow the
evolution of the velocity distribution of the cloud population in gas
moving with velocity +200\kms, typical of the relative velocity offsets
seen in the galaxy profiles; the results are shown in Figure~\ref{monte}.

It can be seen that the peak of the velocity distribution of the cloud
population evolves rapidly to that of the gas in which it is moving; the
width of the velocity distribution becomes progressively narrower but over
a much longer timescale. The resulting velocity distribution is no longer
Gaussian, but can be approximated as a Gaussian distribution plus extended
broad wings (the slight dip at 200\kms\ for the first plotted time
interval should be ignored; it arises only due to the simplicity of the
model in which a cloud whose initial velocity is close to 200\kms\ will
decelerate very slowly). 

The timescale over which the FWHM of the emission line clouds decreases to
that observed in large radio sources (a few hundred \kms) can be used to
test the plausibility of this model. This time interval is $T \approx 5 \times 10^{-7}
t_0$ (the solid line on Figure~\ref{monte}), where $t_0 = 4 n_{\rm c} r_0
/ 3 n_{\rm g}$. Taking $T \sim 3 \times 10^7$\,yr as an appropriate age
for radio sources a few hundred kpc in size, and assuming pre--shock density
ratio $n_{\rm c} / n_{\rm g} \sim  10^4$ with the IGM density decreased a
further factor $\Delta$ by the bow shock, this gives $(r_0/{\rm pc})\Delta
\approx 5$. For $\Delta \approx 40$ (e.g. Clarke \& Burns
1991)\nocite{cla91} the median cloud size would be about about 0.15
parsec; although small, this is certainly plausible given the simplicity
of the assumptions.

\label{lastpage}
\end{document}